\documentclass[runningheads]{llncs}
\usepackage[T1]{fontenc}
\usepackage{graphicx}
\usepackage{booktabs}
\begin{document}
\title{An Exploration of Internal States in Collaborative Problem Solving}
%
%
\author{Sifatul Anindho \and
Videep Venkatesha \and
Mariah Bradford\and Anne M. Cleary\and
Nathaniel Blanchard}
\authorrunning{S. Anindho et al.}
%
\institute{Colorado State University, Fort Collins, CO 80523, USA
\\
\email{\{sifatul.anindho, nathaniel.blanchard\}@colostate.edu}\\
}

\maketitle              
\begin{abstract}
Collaborative problem solving (CPS) is a complex cognitive, social, and emotional process that is increasingly prevalent in educational and professional settings. This study investigates the emotional states of individuals during CPS using a mixed-methods approach. Teams of four first completed a novel CPS task. Immediately after, each individual was placed in an isolated room where they reviewed the video of their group performing the task and self-reported their internal experiences throughout the task. We performed a linguistic analysis of these internal monologues, providing insights into the range of emotions individuals experience during CPS. Our analysis showed distinct patterns in language use, including characteristic unigrams and bigrams, key words and phrases, emotion labels, and semantic similarity between emotion-related words.

\end{abstract}

\section{Introduction}
Collaborative problem solving (CPS) occurs when small teams are required to work together to accomplish a task/goal~\cite{dillenbourg_what_nodate}. Recently, collaboration has been emphasized as an essential skill that is ubiquitous in industry settings. Several recent works have explored the feasibility of automatically parsing these team dynamics \cite{bradford2023automatic,nath2024thoughtshedgehoglinkingdeliberation,vanderhoeven2024multimodal,vanderhoeven2024point,khebour2024commongroundtrackingmultimodal,2024.EDM-long-papers.14}. However, little work has explored individual internal states, e.g., the cognitive and emotional processes that underlie teamwork during team activities. Here, we explore exactly this, specifically focusing on individual reflections of their internal states. 

We hypothesized that combining team task recordings with self-reported individual reflections after the task provides insights into the internal processes that coincide with an individual's actions during the collaborative task. This raises important research questions: What vernacular do individuals use when describing their internal monologues during CPS tasks? What kinds of internal states are the most/least prevalent during collaborative tasks? How do these internal states link to the actions employed during the learning task? 

We examined individuals' internal states as they perform a novel collaborative task. After the task was complete, we collected a spoken retrospective from individuals as they watched a video of their team progressing through the task --- specifically, they self-report what their internal states were moment-to-moment across the task as they watch the video. The collaborative task was a Lego building task in which each participant was given partial information about the Lego structure (see Section \ref{sec:data_collection}). 
Our analysis focuses on identifying patterns in emotion words and themes to help us better understand the emotional experiences that underlie CPS \cite{feyerherm}. Our findings on the distribution of these states also lays the groundwork for future data collections to target distinct internal states.  
\section{Related Work}
CPS has been extensively researched across various domains, including education, psychology, and computer science. To understand the intricacies of team dynamics, researchers have employed multiple methods, such as observational studies, surveys, and physiological measures. For instance, Sun et al. \cite{SUN2020103672} developed a framework for assessing CPS, while Stahl et al. \cite{stahl2014analyzing} examined the roles of individuals in small-group collaboration. Dindar et al. \cite{dindar2022detecting} developed and implemented a new method for detecting shared physiological arousal events (SPAE) during CPS using skin conductance response data. VanderHoeven et al. \cite{vanderhoeven2024multimodal} explored the importance of multimodal features to modeling CPS tasks and outlined various tool sets that can be leveraged to support each of the individual features and their integration. Research on CPS has traditionally focused on externally observable behaviors, with less emphasis on the internal states of individuals engaged in these activities. 

Literature has also shown results on the influence of emotional experiences during learning and problem solving. Graesser et al. \cite{graesser2012autotutor} highlighted the importance of emotions in shaping learning outcomes, demonstrating that emotional states, such as frustration and engagement, play a crucial role in learning \cite{d2012dynamics}. Tyng et al. \cite{tyng2017influences} introduced a basic evolutionary approach to emotion to understand the effects of emotion on learning and memory and the functional roles played by various regions of the brain and their mutual interactions in relation to emotional processing.

Self-reports have been used to assess mental states in various contexts, including learning strategies and approaches. According to Azevedo \cite{azevedo2015defining}, self-reports, in addition to classroom discourse, are the only proven approach that can be used to measure the cognitive, meta cognitive, affective, and motivational constructs of student engagement \cite{gasevic2017detecting}. Beheshitha et al. \cite{beheshitha2016role} made use of existing self-report instruments to measure students' goal orientations and to interpret and triangulate findings obtained through the use of trace data \cite{gavsevic2015let}. Self-report measures have been used to study internal experiences, but primarily with solo activities, leaving a gap in our understanding of internal states during collaborative tasks. Our study addresses this gap by exploring the internal monologues of individuals during CPS tasks.



\section{Methods}
\subsection{Data Collection} \label{sec:data_collection}
\subsubsection{Participants}
This study recruited a total of 32 participants, organized into 8 groups of 4 individuals each. Participants were required to be at least 18 years old, fluent in English and affiliated with Colorado State University as students or staff members. Recruitment efforts were focused within the Computer Science department of Colorado State University, and potential participants were informed of the study through verbal communication.

Table~\ref{tab:dem_data} presents the demographics of our participants. The age of participants ranged from 20 to 32 years. A total of 21 participants identified English as their first language, while others reported first languages including Telugu, Tajik, Hindi, Kannada and Persian. Providing demographic information was voluntary and recorded anonymously. Participants were recruited upon request, and all personal data was de-identified.

\begin{table}[t]
  \centering
  \caption{\textbf{Demographics of Participants}}\label{tab:dem_data}
  \begin{tabular}{llr}
    \hline
    \textbf{Gender} & Male & 21 \\
     & Female & 10 \\
     & Nonbinary & 1 \\
     \hline
    \textbf{Native Language} & English & 21 \\
     & Telugu & 7 \\
     & Tajik & 1 \\
     & Hindi & 1 \\
     & Kannada & 1 \\
     & Persian & 1 \\
     \hline
    \textbf{Age} & 18-24 & 19 \\
     & 25-31 & 12 \\
     & 32+ & 1 \\
    \hline
  \end{tabular}
\end{table}
\begin{figure}
\centering
\includegraphics[width=0.5\textwidth]{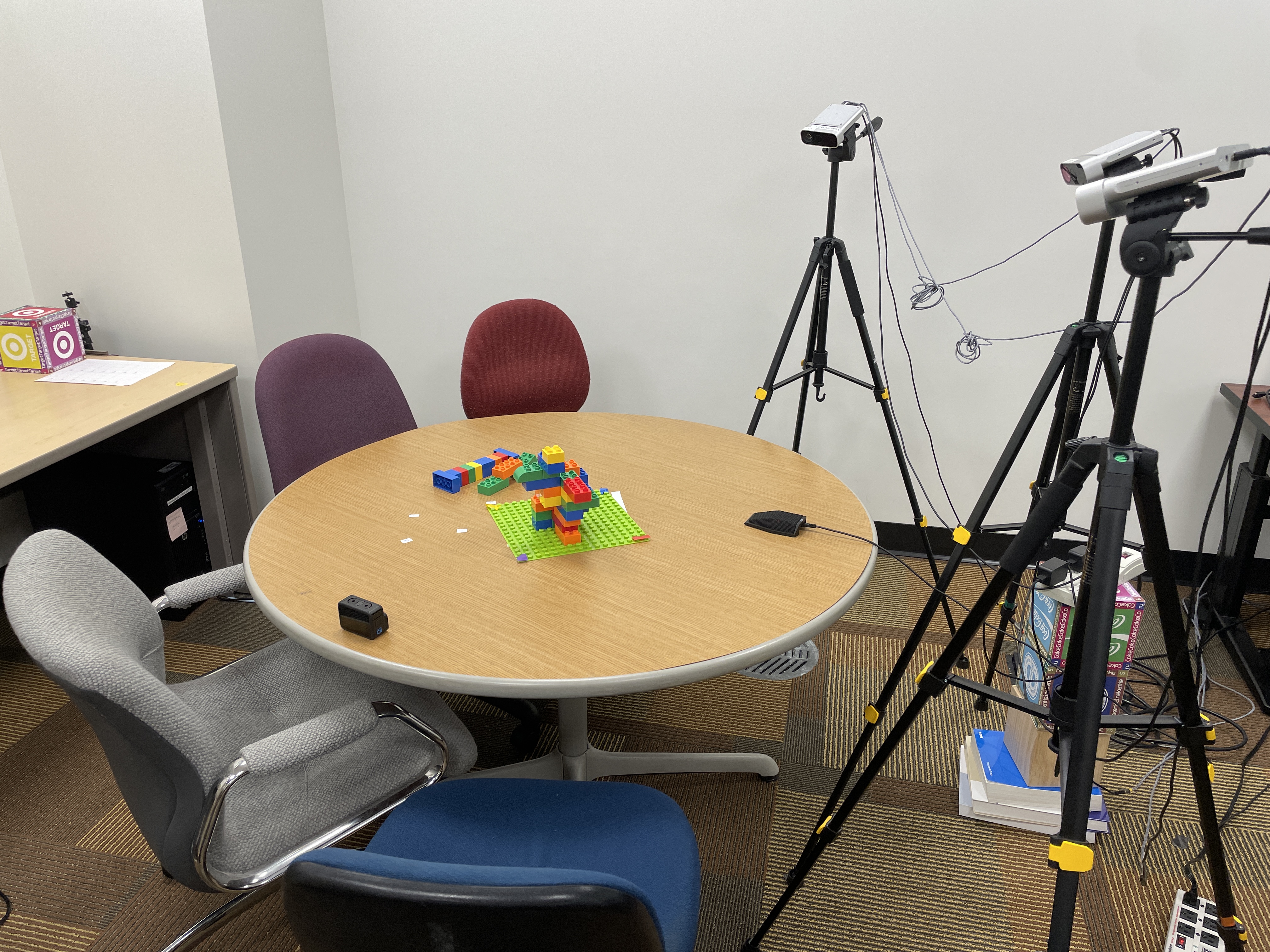}
\caption{Experimental setup for the collaborative Lego building exercise.} \label{fig1}
\end{figure}
To maintain the integrity of group dynamics and task outcomes, participants were allowed to participate in the study multiple times, provided that they were assigned to a different group each time, with at least 3 new members of the group \cite{kerr2004group}. This enabled researchers to collect a more comprehensive data set while minimizing potential biases arising from repeated interactions among familiar group members.
\subsubsection{Experiment Setup}
This study used a collaborative laboratory task paradigm \cite{bahrami2010optimally}. The participants were organized into teams of 4 and completed a two-part situated task involving a Lego building exercise. The task setup (see
Figure~\ref{fig1}) consisted of several key components:
\begin{itemize}
    \item Lego blocks and a building base plate for constructing the Lego structure.
    \item A round table team members collaborated around. 
    \item Azure Kinect Cameras that captured high-quality video and depth information of the task.
    \item A GoPro Camera provided wide-field, high-resolution video for playback during self-report collection.
    \item Skullcandy Hesh Evo headphones used for audio playback. The built-in microphone was used to record participant self-reports.
\end{itemize}

The setup ensured that the participants' faces and hands were clearly visible to the Azure cameras. The Azure cameras were calibrated to ensure clear visuals of the participants during video recording.
\subsubsection{Experiment Procedure}
Before the experiment began, participants reviewed and signed consent forms. Following the consent process, video recordings were initiated on both the Azure Kinect cameras and the GoPro camera to capture RGB, depth video, and audio data. 

The participants were randomly assigned roles: 3 Directors and 1 Builder. Data collectors performed this role assignment to ensure randomization and minimize bias. Directors were provided with tablets displaying  two-dimensional images of the Lego structure generated using Unity-based software. Each image captured the structure from a unique perspective. These images served as the primary reference for the Directors during the task. The task consisted of 2 phases, and the directors received new images of the structure in the second phase. 
Directors were not able to touch blocks and were not allowed to show their images to other participants, requiring them to engage in multimodal collaboration. The builder was only allowed to pick up and place blocks. 

After completion of the task, each participant was placed in a separate room where they received instructions for the self-report section. 
Participants watched the Go-Pro video recording of the first task phase on either a lab computer (Builders) or on their tablets (Directors). As the video played for the participants, they were told to speak aloud their thoughts, feelings and perceptions about themselves, the group members, and the task as they corresponded to the recorded video. Participants were able to pause the video when necessary to elaborate on specific thoughts or feelings, or skip over sections where they did not have anything relevant to talk about. Screen recordings were initiated on each device, capturing participants' self-reports, allowing the participant responses to be mapped back to the original Go-Pro video. 

Participants only reviewed the first phase of the task because both phases of the task were similar in nature and the vernacular of participant monologues was not expected to differ significantly in the second phase \cite{bakeman2011sequential}.

\subsection{Data Preprocessing}
Transcripts of recordings were generated manually by human transcribers using an initial pass with Whisper AI \cite{radford2023robust}. They were separated on the utterance level and saved as a list of segments in JSON format.


Due to poor audio quality, 3 participant self-report recording files were excluded from the analysis, resulting in a sub-sample of 29 participants. This sub-sample consisted of 5 Builders, 8 Director 1s, 8 Director 2s, and 8 Director 3s.

The length of each recording collected from this study is summarized in Table~\ref{tab:sr_data}, along with some descriptive statistics shown in Table \ref{tab:summary_stats}.

\begin{table}[h!]
  \centering
  \caption{\textbf{Length of Recordings from Participant Tasks and Self-Reports}}\label{tab:sr_data}
  \begin{tabular}{lrrrrrr}
    \hline
    \textbf{Group} & \textbf{Task} & \textbf{Builder} & \textbf{Director 1} & \textbf{Director 2} & \textbf{Director 3} \\
    \hline
    1 & 18:08 & 19:34 & 11:56 & 14:31 & 26:59 \\
    2 & 4:04 & 3:03 & 3:55 & 6:03 & 5:01 \\
    3 & 5:44 & 5:42 & 6:42 & 6:13 & 10:43 \\
    4 & 5:56 & 10:57 & 7:37 & 7:28 & 7:06 \\
    5 & 7:15 & 8:09 & 9:41 & 7:18 & 7:14 \\
    6 & 6:07 & *5:49 & 5:53 & 8:58 & 6:22 \\
    7 & 7:45 & *8:24 & 8:26 & 7:46 & 8:18 \\
    8 & 3:06 & *3:12 & 3:21 & 3:15 & 3:12 \\
    \hline
  \end{tabular}
  \vspace{2mm}
  
  \footnotesize{*Data not used for analysis (See 3.2).}
\end{table}

\begin{table}[h!]
  \centering
  \caption{\textbf{Descriptive Statistics for Task Completion and Self-Report (SR) Durations}}\label{tab:summary_stats}
  \setlength{\tabcolsep}{12pt}
  \begin{tabular}{crrr}
    \hline
    \textbf{Category} & \textbf{Task} & \textbf{SR} & \textbf{SR Sub-Sample} \\
    \hline
    Average & 7:15 & 8:05 & 8:19 \\
    SD & 4:39 & 4:53 & 5:02 \\
    Shortest & 3:06 & 3:03 & 3:03 \\
    Longest & 18:08 & 26:59 & 26:59 \\
    \hline
    Total & 58:05 & 4:18:00 & 4:01:23 \\
    \hline
  \end{tabular}
\end{table}



The self-report transcript dataset consisted of 29 transcripts. Table~\ref{tab:summary_statistics} shows the summary statistics of these transcripts.

\begin{table}[h!]
\centering
\caption{Summary Statistics of Self Report (SR) Transcripts}\label{tab:summary_statistics}
\begin{tabular}{l r}
    \toprule
    \textbf{Metric} & \textbf{Value} \\
    \midrule
    Average words per participant & 316 \\
    Standard deviation & 278 \\
    Minimum words per participant & 69 \\
    Maximum words per participant & 1291 \\
    Total number of words & 9153 \\
    \bottomrule
  \end{tabular}
\end{table}
To prepare the transcripts for analysis, several preprocessing steps were applied. The first step involved removing stop words using the NLTK \cite{loper2002nltk} stop-words package, which eliminated common words such as "the," "and," and "is" that do not carry significant meaning \cite{manning1999foundations}. However, to preserve the semantic meaning of negated phrases, we subsequently added back negation words (e.g., "not," "no," "never") to the dataset. This step was necessary to prevent the loss of valuable contextual information that would have been otherwise obscured by the removal of stop words \cite{liu2012survey}. Subsequently, punctuation was removed using the NLTK \cite{loper2002nltk} punctuation package to eliminate non-alphanumeric characters that could interfere with text processing. Additionally, words were normalized to lowercase to standardize all words in a uniform case \cite{manning1999foundations}. Finally, Porter-Stemming was applied to reduce words to their root form \cite{porter1980algorithm}.

\subsection{Data Analysis}

\subsubsection{Frequency Analysis}
We conducted a frequency analysis of unigrams, bigrams and trigrams to identify the most commonly used words and phrases. We visualize uni-grams that occur 30 or more times, bigrams that occur 6 or more time. We also manually extracted key words and phrases that were meaningful in context and determined the most frequent terms. For our analysis, we only consider terms that appear at least 5 times in the data. 

We developed a set of emotion labels based on the distribution of frequent key words and phrases, drawing on work on discrete emotions and their corresponding emotional expressions \cite{ekman1992argument}. Words and phrases were then mapped to these labels. When mapping, we accounted for the the context used by participants.

\subsubsection{Semantic Similarity}
Semantic similarity refers to the degree to which multiple words convey similar meanings or concepts \cite{budanitsky2006evaluating}. Language can be thought of as a multidimensional semantic space, where each word occupies a unique position based on its meaning and relationships to other words \cite{mikolov2013distributed}. To examine the semantic relationships between categories, we used a semantic similarity analysis using the BERT language model \cite{kenton2019bert}. Having a higher score indicates that 2 words are more semantically similar, indicating a commonality in terms of meaning, context, or associations \cite{joulin2016bag}. 

To calculate the semantic similarities, we first generated word embeddings for words using the bert-base-uncased model \cite{geetha2021improving}. This step transformed the words into dense numerical representations that captured their semantic meanings. Next, we calculated the cosine similarity between the word embeddings \cite{rahutomo2012semantic}, which measures the cosine of the angle between 2 vectors. This measure ranges from -1 (dissimilar) to 1 (similar), with 0 indicating orthogonality (no similarity).

We found 2 types of semantic similarities: within-category and between-category, following the methodology used by Hampton and Passanisi \cite{hampton2016intensions} to study individual differences in conceptualization. The within-category represents average semantic similarity between word-pairs that we associate with the same label, and the between-category represents average semantic similarity of word-pairs that we associate with different labels.
\section{Results}
Here, we examine the linguistic details of participants' internal states. First, we examine unigram, bigrams, and trigrams to see which words and phrases were used most frequently (Section \ref{sec:grams}). Because many of these -grams are irrelevant to the internal states of participants during the task, we re-explore the data from a key words and phrases perspective (Section \ref{sec:keywords}). From these, we develop internal-state specific emotion labels (Section \ref{sec:emotions}). Finally, we used a semantic similarity analysis to confirm that these labels capture participants' self-reports as expected and are appropriately distinct. 
\subsection{Unigrams, Bigrams and Trigrams} \label{sec:grams}
\begin{table}[ht]
\centering
\scalebox{0.9}{
\begin{tabular}{lr}
\hline
Unigram & Frequency \\
\hline
like & 175 \\
think & 83 \\
point & 74 \\
kind & 60 \\
thought & 54 \\
yeah & 52 \\
really & 45 \\
right & 44 \\
thinking & 42 \\
would & 40 \\
confused & 38 \\
blocks & 38 \\
side & 35 \\
good & 33 \\
not & 33 \\
trying & 32 \\
one & 31 \\
going & 30 \\
\hline
\\
\end{tabular}
}
\caption{Unigrams Occurring 30 Or More Times in Self-Reported Monologue}
\label{tab:unigrams}
\end{table}


\begin{table}[ht]
\centering
\scalebox{0.9}{
\begin{tabular}{lr}
\hline
Bigram & Frequency \\
\hline
little bit & 15 \\
feel like & 12 \\
like oh & 11 \\
thinking like & 10 \\
point time & 10 \\
make sure & 9 \\
point like & 8 \\
three three & 7 \\
trying figure & 7 \\
little confused & 6 \\
good job & 6 \\
bit confused & 6 \\
look like & 6 \\
\hline
\\
\end{tabular}
}
\caption{Bigrams Occurring 6 Or More Times in Self-Reported Monologue}
\label{tab:bigrams}
\end{table}



Table~\ref{tab:unigrams} shows the most frequent unigrams. Table~\ref{tab:bigrams} shows the same for the bigrams. We also examined the most frequent trigrams --- we found that the phrases "three three matrix", "point time thinking", "done good job", "four four matrix", "point time pretty", "little bit confused", and "thought builder smart" occurred between 3 and 5 times. Of these trigrams, only the "little bit confused" trigram was indicative of an internal state. We reviewed the data and found that in context many of the most frequent words were associated with "filler speech" and were not indicative of internal states.

\begin{table}
\centering
\caption{Key Words and Phrases Occurring 5 or More Times in Self-Reported Monologue}
\label{tab:keyword-frequency}
\scalebox{0.8}{
\begin{tabular}{rlr}
\toprule
Rank & Keyword & Frequency \\
\midrule
1 & confus & 39 \\
2 & good & 28 \\
3 & tri & 18 \\
4 & figure out & 16 \\
5 & right & 12 \\
6 & togeth & 12 \\
7 & think & 11 \\
8 & well & 10 \\
9 & realiz & 9 \\
10 & rememb & 9 \\
11 & smart & 8 \\
12 & easier & 8 \\
13 & easi & 7 \\
14 & happi & 7 \\
15 & wrong & 7 \\
16 & wait & 7 \\
17 & help & 7 \\
18 & figur & 7 \\
19 & understand & 7 \\
20 & thank & 6 \\
21 & correct & 6 \\
22 & confid & 6 \\
23 & glad & 6 \\
24 & make sur & 6 \\
25 & focus & 5 \\
26 & probabl & 5 \\
27 & understood & 5 \\
28 & hope & 5 \\
29 & didn't know & 5 \\
\bottomrule
\end{tabular}
}
\end{table}

\subsection{Key Words and Phrases}\label{sec:keywords}
Our analysis of uni- and bigrams revealed that many of the most frequent were not related to internal states (for example, "like," "think," "little bit," "feel like,"). We manually identified which were related. Table~\ref{tab:keyword-frequency}. shows the identified set of -grams that occurred with a frequency of 5 or more.




A total of 29 keywords met the frequency cutoff, with "confus" being the most frequent (39 occurrences). Other notable keywords included "good", "tri", and "figure out", which occurred 28, 18, and 16 times, respectively. The presence of these keywords suggests that the data contains themes related to confusion, understanding, and problem-solving.

\subsection{Emotion Labels}\label{sec:emotions}
Table \ref{tab:top-5-words-per-label} presents the labels we developed using the key word and phrases, and explicitly indicates which key words and phrases are associated with each label.  

\begin{table}[h!]
  \centering
  \caption{\textbf{Frequent Key Words/Phrases for Each Label}}\label{tab:top-5-words-per-label}
  \small
  \renewcommand{\arraystretch}{1.2} 
  \begin{tabular}{ll}
    \hline
    \textbf{Label} & \textbf{Frequent Key Words and Phrases} \\
    \hline
    Engaged & tri (18), figure out (16), think (11), figur (7), understand (7) \\
    Disengaged & zoning out (4), zoned out (3), slack (1), tuned him out (1), spaced out (1) \\
    Conflicted & wrong (7), conflict (4), problem (3), vagu (3), assum (2) \\
    Confident & realiz (9), confid (6), sure (4), speak up (2), clearli (2) \\
    Reserved & speak up (2), assum (2), quiet (1), spoken up (1), saying noth (1) \\
    Frustrated & stress (2), frustrat (1), crazi (1), stupid (1), hard tim (1) \\
    Optimistic & figure out (16), right (12), easier (8), easi (7), thank (6) \\
    Anxious & worri (3), mistak (2), hyper-fix (1), crazi (1), hard (1) \\
    Disappointed & fault (2), off (2), mistak (2), shoot (1), disappoint (1) \\
    Satisfied & good (28), well (10), easier (8), smart (8), happi (7) \\
    Confused & confus (39), didn't know (5), vagu (3), lost (1), don't know (1) \\
    Surprised & surpris (4), heck (2), oh (2), weird (1), oh god (1) \\
    \hline
  \end{tabular}
\end{table}

\begin{figure}
\centering
\includegraphics[width=0.8\textwidth]{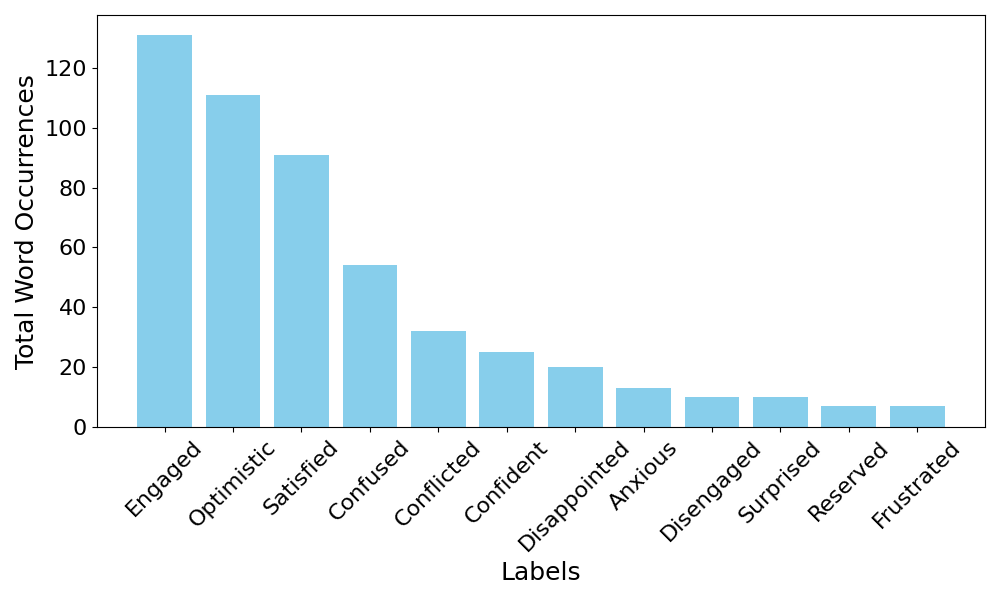}
\caption{The Total Occurrences of Words Associated with each Emotion Theme Label.} \label{fig7}
\end{figure}
\begin{figure}
\centering
\includegraphics[width=0.6\textwidth]
{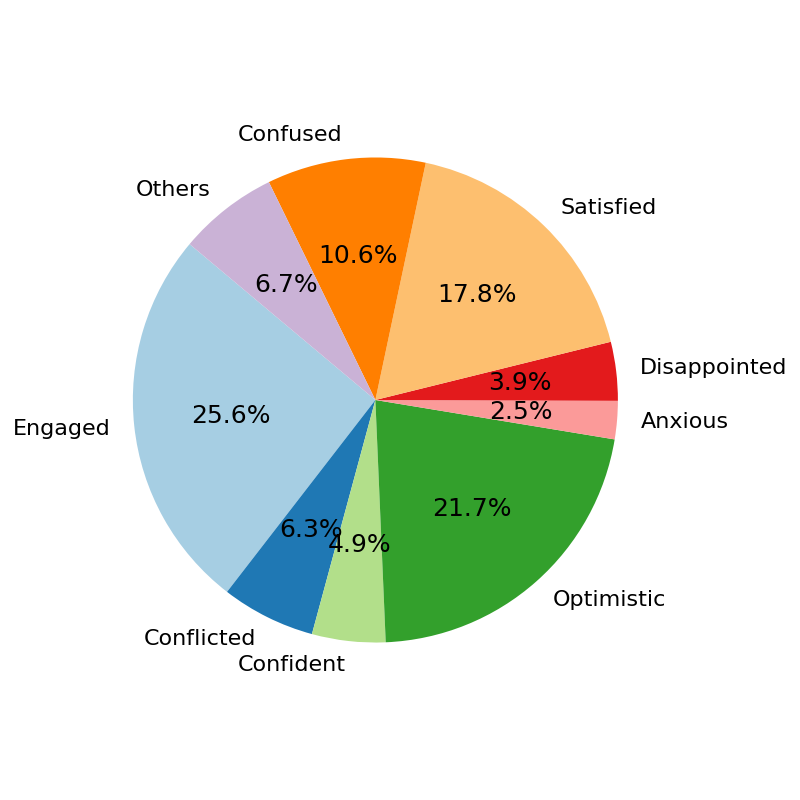}
\caption{The Distribution of the Most Frequent Key Words Categorized by their Respective Emotion Theme Label.} \label{fig8}
\end{figure}

Figure \ref{fig7} and Figure \ref{fig8} present the distribution of internal state labels. "Optimistic", "Engaged", and "Satisfied" labels were the most prevalent, collectively accounting for 333 (61.4\%) of the total word occurrences. In contrast, labels such as "Disengaged", "Reserved", and "Frustrated" were rare --- accounting for only 24 (4.4\%) occurrences. 

\subsection{Semantic Similarity}\label{ref:semantic}

We used semantic similarity to validate that the words within each label were appropriately similar. The within-category semantic similarity analysis reveals that the most semantically similar labels are "Frustrated", "Surprised", and "Conflicted".  In contrast, words in labels such as "Anxious", "Disappointed", and "Confused" exhibited relatively lower semantic similarity, suggesting a more diverse range of semantic meanings within these categories. Table~\ref{tab:within-category-similarity}. shows the results of this analysis. 
\begin{table}[h!]
  \centering
  \caption{\textbf{Within-Category Similarity for Each Emotion Category}}\label{tab:within-category-similarity}
  
  \begin{tabular}{lr}
    \hline
    \textbf{Category} & \textbf{Similarity} \\
    \hline
    Anxious & 0.8679 \\
    Confident & 0.9183 \\
    Conflicted & 0.9297 \\
    Confused & 0.8742 \\
    Disappointed & 0.8731 \\
    Disengaged & 0.8840 \\
    Engaged & 0.8959 \\
    Frustrated & 0.9581 \\
    Optimistic & 0.9146 \\
    Reserved & 0.9367 \\
    Satisfied & 0.9100 \\
    Surprised & 0.9569 \\
    \hline
  \end{tabular}
\end{table}

\begin{figure}
\centering \includegraphics[width=1.0\textwidth]{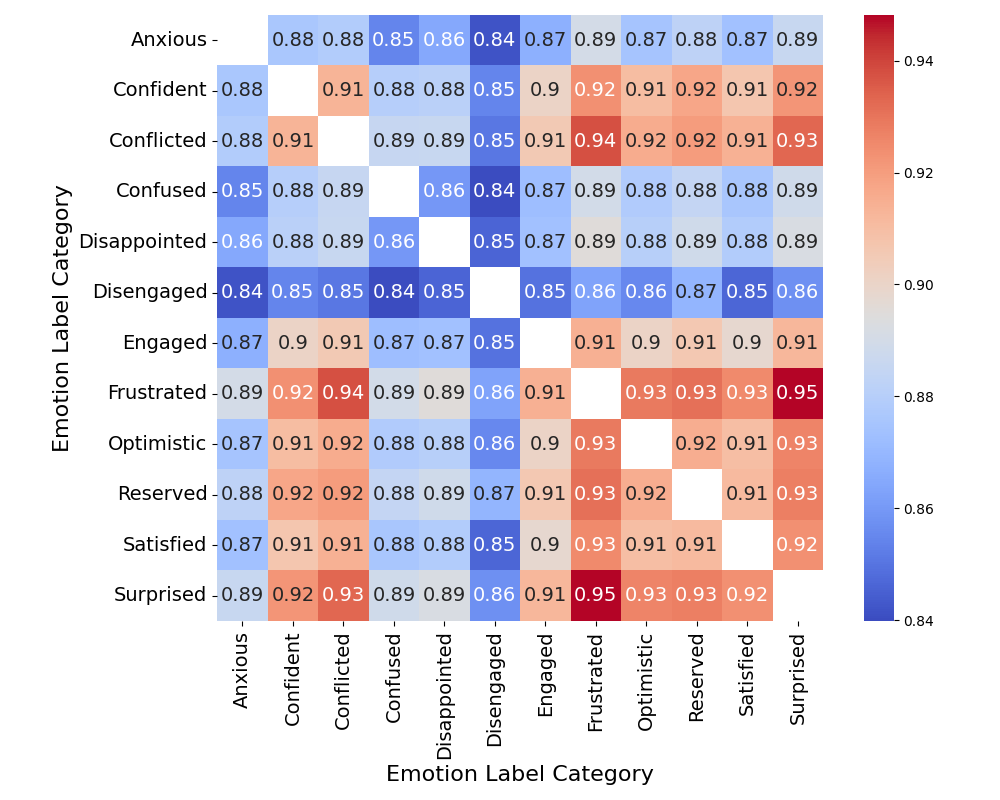}
\caption{Heat-map of the Between-Category Similarity Scores Between Each Pair of Categories.} \label{fig10}
\end{figure}

We also performed a cross-label analysis to investigate if our labels were meaningfully distinct. Figure~\ref{fig10} presents the similarity between categories reveal some patterns in the semantic relationships between emotion labels. The highest similarities between categories was observed between "Frustrated" and "Surprised", "Frustrated" and "Conflicted", and "Surprised" and "Conflicted". The lowest were found between "Disengaged" and "Frustrated", "Disengaged" and "Surprised", and "Anxious" and "Disengaged". These findings provide insight into the semantic structure of emotional experiences during CPS, highlighting the complex relationships between different emotion labels.
\section{Discussion}

Our n-gram analysis shows that frequent phrases were often related to problem solving, communication, and emotions. This suggests that the participants were actively involved in cognitive processes and social interactions during CPS. The presence of phrases of emotional regulation, such as "good job" and "bit confused," implies that the participants were aware of the emotional and cognitive demands of a collaborative task \cite{huang2023social}.

The analysis of key words and phrases highlight the importance of language in shaping emotional experiences during teamwork. Frequent words and phrases such as "figure out," and "understand" indicate a focus on cognitive processing and problem-solving. The presence of words like "good," "well", and "happy" suggests that positive emotional experiences are also prevalent during CPS.

The emotion theme/label analysis reveals that certain labels, such as "Engaged", "Confident", and "Optimistic", were more frequent than others. This suggests that individuals tend to experience positive emotional states during CPS, which is consistent with research on the benefits of teamwork for motivation and engagement \cite{hazel2013team}. In particular, the "Confused" label was relatively frequent. Research has shown that confusion can be beneficial for learning and is indicative of high learner engagement \cite{d2014confusion}. This suggests that confusion may be an important emotional experience during CPS, as it may facilitate deeper understanding and engagement with the task.

We find that words associated with a certain emotion label were similar in semantics. Certain labels, such as "Frustrated", "Conflicted", and "Surprised" had more similarity between words than the rest. This suggests that individuals may experience emotional states that are closely related in meaning, but also distinct from other emotional states. We also see that words in the "Frustrated" and "Surprised" labels were very similar to each other, which could show that the vernacular used to describe these two emotions were closely related. 


\section{Limitations and Future Work}
This study has several limitations that should be acknowledged. First, the size of the participant pool is relatively small, especially for the builder role, which may limit the generalizability of our findings \cite{riener2020addressing}. Second, we make no distinction between the different roles that participants had during the task --- our objective was to capture a holistic perspective of common collaborative internal states, but examining the internal states at the role level may reveal that distinct roles are associated with distinct internal state distributions. 
Third, our reliance on self-reported data may introduce biases, as participants may forget details or not accurately report their emotional experiences after the task is completed \cite{donaldson2002understanding,paulhus2007self}. Finally, the task used in this study is context-specific. Future work needs to confirm that these findings generalize to other CPS contexts \cite{highhouse2009designing}. 

Future work aims to investigate context-specific trends in n-grams and complete sentences, such as internal friction, sudden insight, or conflict. In addition, we plan to utilize the emotional themes developed in this study to design a self-annotation tool that will allow participants to annotate their own task recordings with emotional labels. This will reduce reliance on self-reported data and provide a better understanding of the emotional experiences of individuals during teamwork. 

\section{Conclusion}
In this study, we report the internal states experienced by individuals collaborating with a group. We use frequency and similarity analysis to establish categories of internal states that users experience, as well as the expected distribution of such states. Our study provides a foundation for future research in this area, with important implications for educators, instructional designers, and anyone interested in improving collaborative learning.
\subsubsection{Acknowledgments}
This material is based in part upon work supported by Other Transaction award HR00112490377 from the U.S. Defense Advanced Research Projects Agency (DARPA) Friction for Accountability in Conversational Transactions (FACT) program, and by the National Science Foundation (NSF) under a subcontract to Colorado State University on award DRL 2019805 (Institute for Student-AI Teaming). Approved for public release, distribution unlimited. Views expressed herein do not reflect the policy or position of, the Department of Defense, the National Science Foundation, or the U.S. Government. We thank Nikhil Krishnaswamy, Maniteja Vallala, Shamitha Gowra, Tarun Varma Buddaraju, Sai Kiran Ganesh Kumar, and
Sai Shruthi Garlapati for their support in this study. All errors are the responsibility of the authors.
\bibliographystyle{splncs04}
\bibliography{bib}
\end{document}